\documentclass[11pt,onecolumn,twoside]{IEEEtran}
\usepackage{amsfonts,amsmath,amssymb,epsfig,psfrag,cite,float}
\def\R{{\mathbb{R}}}
\def\var{\mathop{\mathrm{var}}}
\def\yInit{\hat{x}_{\rm init}}
\def\DeltaSNR{\Delta_{\rm SNR}}
\def\ts{\textstyle}
\def\DI{D_{\rm I}}
\def\DII{D_{\rm II}}
\def\Xord{\xi}
\newcommand{\eqlabel}[1]{ \stackrel{(#1)}{=} }
\newcommand{\geqlabel}[1]{ \stackrel{(#1)}{\geq} }
\newcommand{\leqlabel}[1]{ \stackrel{(#1)}{\leq} }
\newcommand{\approxlabel}[1]{ \stackrel{(#1)}{\approx} }
\newtheorem{thm}{Theorem}
\newtheorem{prop}{Proposition}
\newtheorem{conj}{Conjecture}
\newtheorem{lem}{Lemma}
\newcommand{\defeq}{\stackrel{\Delta}{=}}
\newcommand{\Frac}[2]{{{#1}/{#2}}}
\newlength{\widthA}
\newlength{\widthB} 
\floatstyle{ruled}
\newfloat{algorithm}{t}{loa}
\floatname{algorithm}{Algorithm}

\IEEEoverridecommandlockouts

\begin{document}
\title{Concentric Permutation Source Codes%
\thanks{This material is based upon work supported by the National Science
    Foundation under Grant No.\ 0729069\@.  This work was also supported
    in part by a Vietnam Education Foundation Fellowship.}%
\thanks{This work was presented in part at the IEEE International Symposium
    on Information Theory, Seoul, South Korea, June--July 2009.}%
\thanks{The authors are with
        the Department of Electrical Engineering and Computer Science and
        the Research Laboratory of Electronics,
        Massachusetts Institute of Technology,
        Cambridge, MA 02139 USA\@.
        }%
}
\author{Ha Q.\ Nguyen,
        Lav R.\ Varshney, 
        and Vivek K Goyal%
}

\maketitle

\begin{abstract}
Permutation codes are a class of structured vector quantizers with a
computationally-simple encoding procedure based on sorting the scalar
components.  Using a codebook comprising several permutation codes
as subcodes preserves the simplicity of encoding while increasing the
number of rate--distortion operating points, improving the convex
hull of operating points, and increasing design complexity.
We show that when the subcodes are designed
with the same composition, optimization of the codebook reduces to a
lower-dimensional vector quantizer design within a single cone.
Heuristics for reducing design complexity are presented,
including an optimization of the rate allocation in a
shape--gain vector quantizer with gain-dependent wrapped spherical
shape codebook.
\end{abstract}


\section{Introduction}
A permutation source code~\cite{Dunn1965,BergerJW1972} places all codewords
on a single sphere by using the permutations of an \emph{initial codeword}.
The size of the codebook is determined by multiplicities of
repeated entries in the initial codeword,
and the complexity of optimal encoding is low.
In the limit of large vector dimension, an optimal permutation code for
a memoryless source performs as well as entropy-constrained \emph{scalar}
quantization~\cite{Berger1972}.
This could be deemed a disappointment because
the constraint of placing all codewords on a single sphere
does not preclude performance approaching the rate--distortion bound
when coding a memoryless Gaussian source~\cite{Sakrison1968}.
An advantage that remains is that the fixed-rate output of the permutation
source code avoids the possibility of buffer overflow associated with
entropy coding highly nonequiprobable outputs of a quantizer~\cite{Jelinek1968}.

The performance gap between permutation codes and optimal spherical codes,
along with the knowledge that the performance of permutation codes does not
improve monotonically with increasing vector dimension~\cite{GoyalSW2001},
motivates the present paper.
We consider generalizing permutation source codes to have
more than one initial codeword.
While adding very little to the encoding complexity,
this makes the codebook of the vector quantizer (VQ) lie in the
union of concentric spheres rather than in a single sphere.
Our use of multiple spheres is similar to the wrapped spherical
shape--gain vector quantization of Hamkins and Zeger~\cite{HamkinsZ2002};
one of our results, which may be of independent of interest,
is an optimal rate allocation for that technique.
Our use of permutations could be replaced by the action of other
groups to obtain further generalizations~\cite{Slepian1968}.

Design of a permutation source code includes selection of the
multiplicities in the initial codeword; these multiplicities form a \emph{composition}
of the vector dimension~\cite[Ch.~5]{Bona2006}.
The generalization makes the design problem more difficult because there is a
composition associated with each initial codeword.  Our primary focus is on methods
for reducing the design complexity. We demonstrate
the effectiveness of these methods and improvements
over ordinary permutation source codes through simulations.

The use of multiple initial codewords was introduced as
``composite permutation coding'' by Lu \emph{et al.}~\cite{LuCG1986,LuCG1994}
and applied to speech/audio coding by Abe \emph{et al.}~\cite{AbeKN2007}.
These previous works restrict the constituent permutation source codes to
have the same number of codewords, neglect the design of compositions,
and use an iterative VQ design algorithm at the full vector dimension.
In contrast, we allow the compositions to be identical or different,
thus allowing the sizes of subcodes to differ.
In the case of a single, common composition,
we show that a reduced-dimension VQ design problem arises.
For the general case, we provide a rate allocation across subcodes.

The generalization that we study maintains the low $O(n \log n)$ encoding
complexity for vectors of dimension $n$ that permutation source codes achieve.
Vector permutation codes are a different generalization with improved
performance \cite{FinamoreBS2004}.  Their encoding procedure, however,
requires solving the assignment problem in combinatorial
optimization \cite{Kuhn1955} and has complexity $O(n^2 \sqrt{n}\log n)$.

The paper is organized as follows:
We review the attainment of the rate--distortion bound by spherical source codes
and the basic formulation of permutation coding in Section~\ref{sec:background}.
Section~\ref{sec:2} introduces concentric permutation codes
and discusses the difficulty of their optimization.
One simplification that reduces the design complexity---the use of
a single common composition for all initial codewords---is discussed
in Section~\ref{sec:common}.
The use of a common composition obviates the issue of allocating rate
amongst concentric spheres of codewords.
Section~\ref{sec:notcommon} returns to the general case,
with compositions that are not necessarily identical.
We develop fixed- and variable-rate generalizations of
wrapped spherical shape--gain vector quantization
for the purpose of guiding the rate allocation problem.
Concluding comments appear in Section~\ref{sec:conclusions}.

\section{Background}
\label{sec:background}
Let $X \in \R^n$ be a random vector with independent
$\mathcal{N}(0,\sigma^2)$ components.
We wish to approximate $X$ with a \emph{codeword} $\hat{X}$ drawn from a finite \emph{codebook} $\mathcal{C}$.
We want small per-component mean-squared error (MSE) distortion
$D = n^{-1} E[\|X-\hat{X}\|^2]$
when the approximation $\hat{X}$ is represented with $nR$ bits.
In the absence of entropy coding,
this means the codebook has size $2^{nR}$.
For a given codebook, the distortion is minimized when $\hat{X}$ is
the codeword closest to $X$.

\subsection {Spherical Codes}
\label{sec:spherical}
In a \emph{spherical (source) code}, all codewords lie on a single
sphere in $\R^n$.  Nearest-neighbor encoding with such a codebook
partitions $\R^n$ into $2^{nR}$ cells that are (unbounded) convex
cones with apexes at the origin.
In other words, the representations of $X$ and $\alpha X$ are the
same for any scalar $\alpha > 0$.
Thus a spherical code essentially ignores $\| X \|$,
placing all codewords at radius
$$
  E \left[ \| X \| \right] = \frac{\sqrt{2\pi \sigma^2}}{\beta(n/2,1/2)}
        \approx \sigma\sqrt{n - 1/2},
$$
where $\beta(\cdot,\cdot)$ is the beta function, while representing $X / \| X \|$ with $nR$ bits.

Sakrison~\cite{Sakrison1968} first analyzed the performance of spherical
codes for memoryless Gaussian sources.
Following~\cite{Sakrison1968,HamkinsZ2002},
the distortion can be decomposed as
\begin{equation}
  D = \frac{1}{n} E\left[ \left\| \frac{E[\|X\|]}{\|X\|}X - \hat{X} \right\|^2 \right]
            + \frac{1}{n} \var(\|X\|).
\label{eq:Sakrison}
\end{equation}
The first term is the distortion between the projection of $X$ to the
code sphere and its representation on the sphere,
and the second term is the distortion incurred from the projection.
The second term vanishes as $n$ increases even though no bits are spent
to convey the norm of $X$. Placing codewords uniformly at random on the sphere controls the first term
sufficiently for achieving the rate--distortion bound as $n \rightarrow \infty$.

\subsection{Permutation Codes}
\label{sec:permutationcodes}

\subsubsection{Definition and Encoding}
A \emph{permutation code} (PC) is a special spherical code in which all the codewords are related by permutation.
Permutation channel codes were introduced by Slepian~\cite{Slepian1965}
and modified through the duality between source encoding and channel decoding
by Dunn~\cite{Dunn1965}.  They were
then
developed by Berger \emph{et al.}~\cite{BergerJW1972,Berger1972,Berger1982}.

There are two variants of permutation codes:

\emph{Variant I:} Let $\mu_1 > \mu_2 > \cdots > \mu_{K}$ be real numbers, and
let $n_1,n_2,\ldots,n_{K}$ be positive integers with sum equal to $n$
(an \emph{(ordered) composition} of $n$).
The \emph{initial codeword} of the codebook $\mathcal{C}$ has the form
\begin{equation}
    \yInit=(\mathop{{\mu_1,\ldots,\mu_1}}_{\longleftarrow n_1 \longrightarrow},\mathop{{\mu_2,\ldots,\mu_2}}_{\longleftarrow n_2 \longrightarrow},\ldots,\mathop{{\mu_K,\ldots,\mu_K}}_{\longleftarrow n_K \longrightarrow}),
\label{eq:init}
\end{equation}
where each $\mu_i$ appears $n_i$ times.
The codebook is the set of all distinct permutations of $\yInit$.
The number of codewords in $\mathcal{C}$ is thus given by the multinomial coefficient
\begin{equation}
M=\frac{n!}{n_1 ! \, n_2 !\cdots n_{K} !}.
\label{eq:multinomial}
\end{equation}

The permutation structure of the codebook enables low-complexity
nearest-neighbor encoding~\cite{BergerJW1972}: map $X$ to the codeword
$\hat{X}$ whose components have the same order as $X$;
in other words, replace the $n_1$ largest components of $X$ with $\mu_1$,
the $n_2$ next-largest components of $X$ with $\mu_2$, and so on.

\emph{Variant II:} The initial codeword $\yInit$ still has the form
(\ref{eq:init}), but now all its entries are nonnegative;
i.e., $\mu_1>\mu_2 \mbox{$>\cdots$}>\mu_K\geq 0$.
The codebook now consists of all possible permutations of $\yInit$ in which each nonzero component is possibly negated.
The number of codewords is thus given by
\begin{equation}
M=2^{h}\cdot \frac{n!}{n_1 ! \, n_2 !\ldots n_{K} !},
\end{equation}
where $h$ is the number of positive components in $\yInit$.
Optimal encoding is again simple~\cite{BergerJW1972}: map $X$ to the codeword
$\hat{X}$ whose components have the same order in absolute value and match the signs of corresponding components of $X$.

Since the complexity of sorting is $O(n \log n)$ operations,
the encoding complexity
is much lower than with an unstructured
VQ
and only $O(\log n)$ times higher than scalar quantization.

\subsubsection{Performance and Optimization}
For i.i.d.\ sources, each codeword is chosen with equal probability. Consequently, there is no improvement from entropy coding and the per-letter rate is simply $R=n^{-1}\log M$.

Let $\xi_1\geq\xi_2\geq\cdots\geq\xi_n$ denote the order statistics of random vector $X=\left(X_1,\ldots,X_n\right)$, and $\eta_1\geq\eta_2\geq\cdots\geq\eta_n$ denote the order statistics of random vector $|X|\defeq\left(|X_1|,\ldots,|X_n|\right)$.%
\footnote{Because of the
convention $\mu_i > \mu_{i+1}$ established by Berger \emph{et al.}~\cite{BergerJW1972}, it is natural to
index the order statistics in descending order as shown, which is opposite to
the ascending convention in the order statistics literature~\cite{DavidN2003}.}
With these notations and an initial codeword given by (\ref{eq:init}),
the per-letter distortion of optimally-encoded Variant~I and Variant~II codes
can be deduced simply by realizing which order statistics are mapped to each
element of $\yInit$:
\begin{eqnarray}
\DI & = & n^{-1}E\left[{\ts\sum_{i=1}^{K}\sum_{\ell \in I_i}\left(\xi_\ell-\mu_i\right)^2}\right] \qquad \mbox{and} \label{eq:distort} \\
\DII & = & n^{-1}E\left[{\ts\sum_{i=1}^{K}\sum_{\ell \in I_i}\left(\eta_\ell-\mu_i\right)^2}\right] , \label{eq:distort2}
\end{eqnarray}
where $I_i$s are the groups of indices generated by the composition, i.e.,
\[
I_1 = \{1,2, \ldots ,n_1 \} \mbox{,} \quad I_i = \left\{\left({\textstyle\sum_{m=1}^{i-1}n_m}\right)+1,\ldots, \left({\textstyle\sum_{m=1}^{i}n_m}\right)\right\},\ i\geq 2.
\]

Given a composition $(n_1,n_2,\ldots,n_K)$, minimization of $\DI$ or $\DII$ can be done separately for each $\mu_i$, yielding optimal values
\begin{equation}
\mu_i=n_i^{-1}\ts\sum_{\ell\in I_i}E\left[\xi_\ell\right], \quad \mbox{for Variant I, and}\label{eq:optInit1}
\end{equation}
\begin{equation}
\mu_i=n_i^{-1}\ts\sum_{\ell\in I_i}E\left[\eta_\ell\right], \quad \mbox{for Variant II.}\label{eq:optInit2}
\end{equation}
Overall minimization of $\DI$ or $\DII$ over the choice of $K$, $\{n_i\}_{i=1}^K$, and $\{\mu_i\}_{i=1}^K$ subject to a rate constraint is difficult because of the integer constraint of the composition.

The analysis of \cite{Berger1972} shows that as $n$ grows large,
the composition can be designed to give performance equal to
optimal entropy-constrained scalar quantization (ECSQ) of $X$.
Heuristically, it seems that for large block lengths,
PCs suffer because there are
too many permutations ($n^{-1} \log_2 n!$ grows)
and the vanishing fraction that are chosen to meet a rate constraint
do not form a good code.
The technique we study in this paper is for moderate values of $n$,
for which the second term of (\ref{eq:Sakrison}) is not negligible;
thus, it is not adequate to place all codewords on a single sphere.

\section{Permutation Codes with Multiple Initial Codewords}
\label{sec:2}

In this paper, we generalize ordinary PCs by allowing multiple
initial codewords. 
The resulting codebook is contained in a set of concentric spheres.

\subsection{Basic Construction}

Let $J$ be a positive integer.  We will define a \emph{concentric permutation
(source) code} (CPC) with $J$ initial codewords.  This is equivalent to having a
codebook that is the union of $J$ PCs.
Each notation from Section~\ref{sec:permutationcodes}
is extended with a superscript or subscript $j \in \{1,\,2,\ldots,J\}$
that indexes the constituent PC\@.  Thus,
$\mathcal{C}_j$ is the subcodebook of full codebook
$\mathcal{C} = \cup_{j=1}^J \mathcal{C}_j$
consisting of all $M_j$ distinct permutations of initial vector
\begin{equation}
\yInit^{j} = \left(\mu^{j}_{1}, \ldots ,\mu^{j}_{1},\ldots , \mu^{j}_{K_j},\ldots , \mu^{j}_{K_j}
 \right),
\end{equation}
where each $\mu^j_i$ appears $n^j_i$ times,
$\mu^j_1 > \mu^j_2 > \cdots > \mu^j_{K_j}$
(all of which are nonnegative for Variant~II),
and $\sum_{i=1}^{K_j} n^j_i = n$.
Also, $\{I^j_i\}_{i=1}^{K_j}$ are sets of indices generated by the $j$th
composition.

\begin{prop}
Nearest-neighbor encoding of $X$ with codebook $\mathcal{C}$
can be accomplished with the following procedure:
\begin{enumerate}
\item For each $j$, find $\hat{X}_j \in \mathcal{C}_j$ whose components have the same order as $X$.
\item Encode $X$ with $\hat{X}$, the nearest codeword amongst $\{\hat{X}_j\}_{j=1}^J$.
\end{enumerate}
\end{prop}

\begin{IEEEproof}
Suppose $X'\in\mathcal{C}$ is an arbitrary codeword. Since $\mathcal{C}=\cup_{j=1}^J\mathcal{C}_j$, there must exist $j_0\in\{1,2,\ldots,J\}$ such that $X'\in\mathcal{C}_{j_0}$. We have
\[
\|X-\hat{X}\|
  \leqlabel{a} \|X-\hat{X}_{j_0}\|
  \leqlabel{b} \|X-X'\| ,
\]
where (a) follows from the second step of the algorithm, and (b) follows from the first step and the optimality of the encoding for ordinary PCs.
\end{IEEEproof}

The first step of the algorithm requires $O(n\log n) + O(Jn)$ operations (sorting components of $X$ and reordering each $\yInit^j$ according to the index matrix obtained from the sorting); the second step requires $O(Jn)$ operations. The total complexity of encoding is therefore $O(n\log n)$, provided that we keep $J = O(\log n)$. In fact, in this rough accounting, the encoding with $J = O(\log n)$ is as cheap as the encoding for ordinary PCs.

For i.i.d.\ sources, codewords within a subcodebook are approximately equally likely to be chosen, but codewords in different subcodebooks may have very different probabilities. Using entropy coding yields
\begin{equation}
R \approx n^{-1} \left[H\left(\{p_j\}_{j=1}^J\right)+ {\ts\sum_{j=1}^{J}p_j\log M_j}\right], \label{eq:rate}
\end{equation}
where $H(\cdot)$ denotes the entropy of a distribution, $p_j$ is the probability of choosing subcodebook $\mathcal{C}_j$, and $M_j$ is the number of codewords in $\mathcal{C}_j$.
Note that (\ref{eq:rate}) is suggestive of a two-stage encoding scheme
with a variable-rate code for the index of the chosen subcodebook and
a fixed-rate code for the index of the chosen codeword within the subcodebook.
Without entropy coding, the rate is
\begin{equation}
\label{eq:fixed-rateRate}
R = n^{-1} \log\left(\ts\sum_{j=1}^J M_j\right).
\end{equation}

The per-letter distortion for Variant~I codes is now given by
\begin{equation}
D  = n^{-1} E\left[\min_{1\leq j \leq J}\|X-\hat{X}_j\|^2 \right]
   = n^{-1} E\left[\min_{1\leq j \leq J} { \ts\sum_{i=1}^{K_j}\sum_{\ell \in I^j_i}\left(\xi_\ell-\mu^j_i\right)^2}\right], \label{eq:distortion}
\end{equation}
where (\ref{eq:distortion}) is obtained by rearranging the components of $X$ and $\hat{X}_j$ in descending order.
The distortion for Variant~II codes has the same form as (\ref{eq:distortion}) with $\{\xi_\ell\}$ replaced by $\{\eta_\ell\}$.

\subsection{Optimization}

In general, finding the best ordinary PC requires an
exhaustive search over all compositions of $n$.
(Assuming a precomputation of all the order statistic means,
the computation of the distortion
for a given composition through either (\ref{eq:distort}) or (\ref{eq:distort2}) is simple~\cite{BergerJW1972}.)
The search space can be reduced for certain distributions of $X$
using~\cite[Thm.~3]{BergerJW1972}, but seeking the optimal code still
quickly becomes intractable as $n$ increases.

Our generalization makes the design problem considerably more difficult.
Not only do we need $J$ compositions, but the distortion for a given
composition is not as easy to compute.  Because of the minimization over $j$
in (\ref{eq:distortion}), we lack a simple expression for $\mu^j_i$s
in terms of the composition and the order statistic means as given in (\ref{eq:optInit1}).  The relevant means
are of conditional order statistics, conditioned on which subcodebook is
selected; this depends on all $J$ compositions.

In the remainder of the paper, we consider two ways to reduce the
design complexity.  In Section~\ref{sec:common}, we fix all subcodebooks
to have a common composition.  Along with reducing the design space,
this restriction induces a structure in the full codebook that enables
the joint design of $\{ \mu_i^j \}_{j=1}^J$ for any $i$.
In Section~\ref{sec:notcommon}, we take a brief detour into the optimal
rate allocations in a wrapped spherical shape--gain vector quantizer
with gain-dependent shape codebook.  We use these rate allocations to
pick the sizes of subcodebooks $\{\mathcal{C}_j\}_{j=1}^J$.

The simplifications presented here still leave high design complexity for large $n$.  Thus, some simulations use complexity-reducing heuristics including our conjecture that an analogue to~\cite[Thm.~3]{BergerJW1972} holds. Since our numerical designs are not provably optimal, the improvements from allowing multiple initial codewords could be somewhat larger than we demonstrate.

\section{Design with Common Composition}
\label{sec:common}
In this section, assume that the $J$ compositions are identical, i.e.,
the $n_i^j$s have no dependence on $j$.
The subcodebook sizes are also equal, and dropping unnecessary
sub- and superscripts we write the common composition as $\{n_i\}_{i=1}^K$
and the size of a single subcodebook as $M$.

\subsection{Common Compositions Give Common Conic Partitions}
The Voronoi regions of the code now have a special geometric structure.
Recall that any spherical code partitions $\R^n$ into (unbounded) convex cones.
Having a common composition implies that each subcodebook induces the same
conic Voronoi structure on $\R^n$. The full code divides each of the $M$ cones
into $J$ Voronoi regions.

The following theorem precisely maps the encoding of a CPC to a vector quantization problem.
For compositions other than $(1,1,\ldots,1)$,
the VQ design problem is in a dimension strictly lower than $n$.

\begin{thm}\label{thm:VQ}
For fixed common composition $(n_1,n_2,\ldots,n_K)$, the initial codewords\\
$\{(\mu^{j}_{1}, \ldots ,\mu^{j}_{1},\ldots , \mu^{j}_{K},\ldots , \mu^{j}_{K})\}_{j=1}^J$ of a Variant~I CPC are optimal if and only if $\{\mathbf{\mu}^1,\ldots ,\mathbf{\mu}^J \}$ are representation points of the optimal $J$-point vector quantization of
$\overline{\mathbf{\xi}} \in \R^K$, where
\[
    \mu^j=\left(\sqrt{n_1}\,\mu^j_1,\,
                \sqrt{n_2}\,\mu^j_2,\,
                \ldots,\,
                \sqrt{n_K}\,\mu^j_K\right), \qquad 1 \leq j \leq J ,
\]
\[
    \overline{\xi}=\left(\frac{1}{\sqrt{n_1}}{\ts\sum_{\ell\in I_1}\xi_\ell},\,
                         \frac{1}{\sqrt{n_2}}{\ts\sum_{\ell\in I_2}\xi_\ell},\,
                         \ldots,\,
                         \frac{1}{\sqrt{n_K}}{\ts\sum_{\ell\in I_K}\xi_\ell}\right) .
\]
\end{thm}
\begin{IEEEproof}
Rewrite the distortion as follows:
\begin{eqnarray}
nD &=& E\left[\min_{1\leq j\leq J}\sum_{i=1}^K \sum_{\ell\in I_i}{(\xi_\ell-\mu^j_i)}^2\right] = E\left[\min_{1\leq j\leq J}\sum_{i=1}^K \left( \sum_{\ell\in I_i}(\xi_\ell)^2-2\mu^j_i\sum_{\ell\in I_i}\xi_\ell+ n_i(\mu^j_i)^2 \right)\right] \nonumber\\
   &=& E\left[\min_{1\leq j\leq J}\sum^K_{i=1}\left(\frac{1}{\sqrt{n_i}}\sum_{\ell\in I_i}\xi_\ell-\sqrt{n_i}\mu^j_i\right)^2\right] + E\left[\sum_{i=1}^K\sum_{\ell\in I_i}(\xi_\ell)^2 \right]-E\left[\sum_{i=1}^K\left(\frac{1}{\sqrt{n_i}}\sum_{\ell\in I_i}\xi_\ell\right)^2\right]\nonumber\\
   &=& E\left[\min_{1\leq j \leq J}\|\overline{\xi}-\mu^j\|^2\right]+ E\left[\|X\|^2 \right] -E\left[\sum_{i=1}^K\left(\frac{1}{\sqrt{n_i}}\sum_{\ell\in I_i}\xi_\ell\right)^2\right].\label{eq:totaldistortion}
\end{eqnarray}
Since the second and third terms of (\ref{eq:totaldistortion}) do not depend on $\{\yInit^{j}\}_{j=1}^J$, minimizing $D$ is equivalent to minimizing the first term of (\ref{eq:totaldistortion}). By definition of a $K$-dimensional VQ, that term is minimized if and only if $\{\mathbf{\mu}^1,\ldots ,\mathbf{\mu}^J \}$ are optimal representation points of the $J$-point VQ of random vector $\overline{\mathbf{\xi}}$, completing the proof.
\end{IEEEproof}

For any fixed composition, one can implement the $J$-point
VQ design inspired by Theorem~\ref{thm:VQ}, using the Lloyd-Max algorithm~\cite{Lloyd1957,Max1960},
to obtain
$\{\mathbf{\mu}^1,\ldots ,\mathbf{\mu}^J \} \subset \R^K$
and then apply the mapping stated in the theorem to obtain the
$J$ desired initial codewords in $\R^n$.
Theorem~\ref{thm:VQ} can be trivially extended for Variant~II codes by simply replacing $\{\xi_\ell\}$ with $\{\eta_\ell\}$.

\begin{figure}
 \centering
  \includegraphics[width=\widthA]{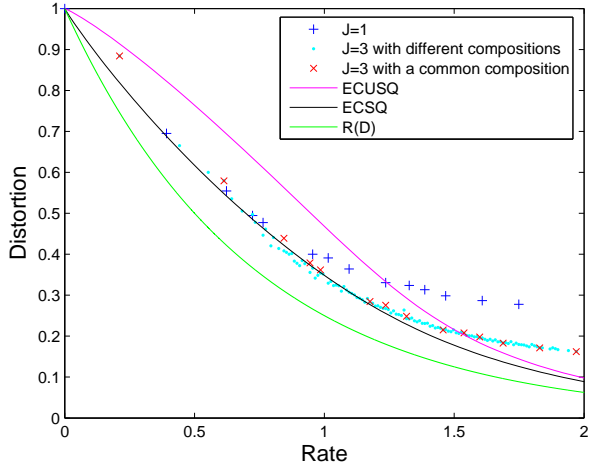}
  \caption{\label{fig:RDexhaustive} Rate--distortion
    performance for variable-rate coding of i.i.d.\
    $\mathcal{N}(0,1)$ source with block length $n=7$.
    Ordinary Variant~I PCs ($J=1$) are compared with
    CPCs with $J=3$.
    Codes with common compositions are designed according to Theorem~\ref{thm:VQ}.
    Codes with different compositions are designed with heuristic selection of
    compositions guided by Conjecture~\ref{conjecture2} and Algorithm~\ref{alg:lloyd}.  For clarity, amongst approximately-equal rates, only operational points with the lowest distortion are plotted.}
\end{figure}

Figure~\ref{fig:RDexhaustive} compares the performance of an ordinary Variant~I PC
($J=1$) with variable-rate CPCs with $J=3$ initial vectors. For a given composition, the distortion of the optimal ordinary PC is computed using (\ref{eq:optInit1}) and variances of the order statistics (see~\cite[Eq.~(13)]{BergerJW1972}), whereas that of the optimal CPC is estimated empirically from 500\,000 samples generated according to the $\mathcal{N}(0,1)$ distribution.
Figure~\ref{fig:RDexhaustive} and several subsequent figures include for comparison the
rate--distortion bound and the performances of two types of entropy-constrained
scalar quantization:
uniform thresholds with uniform codewords (labeled ECUSQ)
and
uniform thresholds with optimal codewords (labeled ECSQ).
At all rates, the latter is a very close approximation to optimal ECSQ;
in particular, it has optimal rate--distortion slope at rate zero
\cite{MarcoN2006}.

\subsection{Optimization of Composition}
Although the optimization of compositions is not easy even for ordinary PCs, for a certain class of distributions, there is a useful necessary condition for the optimal composition~\cite[Thm.~3]{BergerJW1972}. The following conjecture is an analogue of that condition.
\begin{conj}\label{conjecture1}
    Suppose that $J>1$ and that $E[\eta_\ell]$ is a convex function of $\ell$, i.e.
    \begin{equation}
        E\left[\eta_{\ell+2}\right]-2 \,E\left[\eta_{\ell+1}\right]+E\left[\eta_{\ell}\right]\geq 0, \qquad  1\leq \ell \leq n-2.
    \end{equation}
    Then the optimum $n_i$ for Variant~II CPCs
    increases monotonically with $i$.
\end{conj}

The convexity of $E[\eta_\ell]$ holds for a large class of source distributions
(see~\cite[Thm.~4]{BergerJW1972}), including Gaussian ones.
Conjecture~\ref{conjecture1} greatly reduces the search space for optimal
compositions for such sources.

The conjecture is proven if one can show that the distortion associated with the composition $(n_1,\ldots , n_m,n_{m+1},\ldots , n_K)$, where $n_m > n_{m+1}$, can be decreased by reversing the roles of $n_m$ and $n_{m+1}$. As a plausibility argument for the conjecture, we will show that the reversing has the desired property when an additional constraint is imposed on the codewords. With the composition fixed, let
\begin{equation} \label{eq:rv}
        \zeta\defeq \frac{1}{r}\sum_{L+1}^{L+r}\eta_\ell-\frac{2}{q-r}\sum_{L+r+1}^{L+q}\eta_\ell+\frac{1}{r}\sum_{L+q+1}^{L+q+r}\eta_\ell,        
\end{equation}
where $L=n_1+n_2+\ldots + n_{m-1}$.
The convexity of $E[\eta_\ell]$ implies the nonnegativity of $E[\zeta]$
(see \cite[Thm.~2]{BergerJW1972}).
Using the total expectation theorem, $E[\zeta]$ can be written as the difference of two nonnegative terms,
\[
  \overline{\zeta}_{+}\defeq \mbox{Pr}(\zeta\geq 0)E[\zeta \mid \zeta\geq 0]
  \qquad
  \mbox{and}
  \qquad
  \overline{\zeta}_{-}\defeq -\mbox{Pr}(\zeta < 0)E[\zeta \mid \zeta < 0].
\]
Since $E[\zeta]\geq 0$ and probabilities are nonnegative, it is clear that $\overline{\zeta}_{+}\geq \overline{\zeta}_{-}$. Therefore,
the following set is non-empty:
\begin{equation}
\Omega_m = \left\{\left\{\mu^j_i\right\}_{i,j} \mbox{ s.t. } \frac{\min_{j}{(\mu^j_m-\mu^j_{m+1})}}{\max_{j}{(\mu^j_m-\mu^j_{m+1})}} \geq \frac{\overline{\zeta}_{-}}{\overline{\zeta}_{+}}\right\}.\label{eq:constraint}
\end{equation}
With the notations above, we are now ready to state the proposition.
If the restriction of the codewords were known to not preclude optimality,
then Conjecture~\ref{conjecture1} would be proven.
\begin{prop}\label{thm:reverse}
    Suppose that $J>1$ and $E[\eta_\ell]$ is a convex function of $\ell$. If $n_m > n_{m+1}$ for some $m$, and the constraint $\Omega_m$ given in (\ref{eq:constraint}) is imposed on the codewords,
    then the distortion associated with the composition $(n_1,\ldots , n_m,n_{m+1},\ldots , n_K)$ can be decreased by reversing the roles of $n_m$ and $n_{m+1}$.
\end{prop}
\begin{IEEEproof}
See Appendix~\ref{app:proof}.
\end{IEEEproof}

A straightforward extension of Conjecture~\ref{conjecture1} for Variant~I codes is the following:
\begin{conj}\label{conjecture2}
Suppose that $J>1$, and that $E[\xi_\ell]$ is convex over $\mathcal{S}_1\triangleq\{1,2,\ldots,\lfloor K/2 \rfloor\}$ and concave over $\mathcal{S}_2\triangleq\{\lfloor K/2 \rfloor + 1, \lfloor K/2 \rfloor + 2,\ldots,K\}$. Then the optimum $n_i$ for Variant~I CPCs
increases monotonically with $i\in\mathcal{S}_1$ and decreases monotonically with $i\in\mathcal{S}_2$.
\end{conj}

The convexity of $E[\xi_\ell]$ holds for a large class of source distributions
(see~\cite[Thm.~5]{BergerJW1972}).  We will later restrict the compositions, while doing simulations for Variant~I codes and Gaussian sources, to satisfy Conjecture~\ref{conjecture2}.

\section{Design with Different Compositions}
\label{sec:notcommon}
Suppose now that the compositionss of subcodebooks can be different.
The Voronoi partitioning of $\R^n$ is much more complicated,
lacking the separability discussed in the previous section.%
\footnote{For a related two-dimensional visualization,
compare \cite[Fig.~3]{SwaszekT1982} against \cite[Figs.~7--13]{SwaszekT1982}.}
Furthermore, the apparent design complexity for the compositions is
increased greatly to equal the number of compositions raised to the $J$th power, namely $2^{J(n-1)}$.

In this section we first outline an algorithm for local optimization of initial vectors with all the compositions fixed. Then we address a portion of the composition design problem which is the sizing of the subcodebooks. For this, we extend the high-resolution analysis of~\cite{HamkinsZ2002}. For brevity, we limit our discussion to Variant~I CPCs; Variant~II could be generalized similarly.

\subsection{Local Optimization of Initial Vectors}
\label{sec:local}
Let $\Xord=(\xi_1,\xi_2,\ldots,\xi_n)$ denote the ordered vector of $X$. Given $J$ initial codewords $\{\yInit^j\}_{j=1}^J$,
for each $j$,
let $R_j \subset\mathbb{R}^n$ denote the quantization region of $\Xord$ corresponding to codeword $\yInit^j$, and let $E_j[\cdot]$ denote the expectation conditioned on $\Xord \in R_j$.
If $R_j$ is fixed, consider the distortion conditioned on
$\Xord \in R_j$
\begin{equation}
D_j=n^{-1} E\left[\ts\sum_{i=1}^{K_j}\sum_{\ell \in I^j_i}\left(\xi_\ell-\mu^j_i\right)^2 \mid \Xord \in R_j\right].
\end{equation}
By extension of an argument in~\cite{BergerJW1972}, $D_j$ is minimized with
\begin{equation}
\mu^j_i=\frac{1}{n^j_i}\ts\sum_{\ell\in I^j_i}E_j[\xi_\ell], \qquad 1 \leq i\leq K_j . \label{eq:mu}
\end{equation}
For a given set $\{R_j\}_{j=1}^J$, since the total distortion is determined by
\[
D=\ts\sum_{j=1}^J\mbox{Pr}(\Xord \in R_j)D_j ,
\]
it will decrease if $\mu^j_i$s are set to the new values given by (\ref{eq:mu}) for all $1\leq j\leq J$ and for all $1 \leq i\leq K_j$.

From the above analysis, a Lloyd algorithm can be developed to design initial codewords as given in Algorithm~\ref{alg:lloyd}. This algorithm is similar to the algorithm in~\cite{LuCG1986}, but here the compositions can be arbitrary.
Algorithm~\ref{alg:lloyd} was used to produce the operating points shown in
Figure~\ref{fig:RDexhaustive} for CPCs with different compositions in which the distortion of a locally-optimal code was computed empirically from 500\,000 samples generated according to $\mathcal{N}(0,1)$ distribution. We can see through the figure that common compositions can produce almost the
same distortion as possibly-different compositions for the same rate.
However, allowing the compositions to be different yields many more rates.
The number of rates is explored in Appendix~\ref{app:rate-points}.

\begin{algorithm}
 \caption{Lloyd Algorithm for Initial Codeword Optimization from Given Composition}
 \label{alg:lloyd}
 \begin{tabular}{p{6.5in}}
    \begin{enumerate}
        \item Order vector $X$ to get $\Xord$
        \item Choose an arbitrary initial set of $J$ representation vectors $\yInit^1, \yInit^2, \ldots,\yInit^J$.
        \item For each $j$, determine the corresponding quantization region $R_j$ of $\Xord$.
        \item For each $j$, set $\yInit^j$ to the new value given by (\ref{eq:mu}).
        \item Repeat steps 3 and 4 until further improvement in MSE is negligible.
    \end{enumerate}\\
 \end{tabular}
\end{algorithm}

\subsection{Wrapped Spherical Shape--Gain Vector Quantization}
\label{sec:wrapped}
Hamkins and Zeger~\cite{HamkinsZ2002} introduced a
type of spherical code for $\R^n$ where a lattice in $\R^{n-1}$ is ``wrapped''
around the code sphere.
They applied the wrapped spherical code (WSC)
to the shape component in a shape--gain vector quantizer.

We generalize this construction to allow the size of the shape codebook to
depend on the gain. Along this line of thinking,
Hamkins~\cite[pp.~102--104]{Hamkins1996} provided an algorithm
to optimize the number of codewords on each sphere. However, neither analytic nor experimental improvement was demonstrated. In contrast, our approach based on high-resolution optimization gives an explicit expression for the improvement in signal-to-noise ratio (SNR). While our results may be of independent interest, our present purpose is to guide the selection of $\{M_j\}_{j=1}^J$ in CPCs.

A \emph{shape--gain} vector quantizer (VQ) decomposes a source vector $X$ into a \emph{gain} $g=\|X\|$ and a \emph{shape} $S=X/g$, which are quantized to $\hat{g}$ and $\hat{S}$, respectively, and the approximation is $\hat{X}=\hat{g}\cdot\hat{S}$.
We optimize here a wrapped spherical VQ with gain-dependent shape codebook. The gain codebook, $\{\hat{g}_1,\hat{g}_2,\ldots,\hat{g}_J\}$, is optimized for the gain pdf, e.g., using the scalar Lloyd-Max algorithm~\cite{Lloyd1957,Max1960}. For each gain codeword $\hat{g}_j$, a shape subcodebook is generated by wrapping the sphere packing $\Lambda \subset \R^{n-1}$ on to the unit sphere in $\R^n$. The same $\Lambda$ is used for each $j$, but the density (or scaling) of the packing may vary with $j$. Thus the normalized second moment $G(\Lambda)$ applies for each $j$ while minimum distance $d_{\Lambda}^j$ depends on the quantized gain $\hat{g}_j$. We denote such a sphere packing as $(\Lambda,d_{\Lambda}^j)$.

The per-letter MSE distortion will be
\begin{eqnarray*}
D &=& n^{-1} E\left[\|X-\hat{g}\,\hat{S}\|^2\right]\\
  &=& n^{-1} E\left[\|X-\hat{g}\,S\|^2\right]
      +2n^{-1} E\left[({X}-\hat{g}\,S)^{T}(\hat{g}\,S-\hat{g}\,\hat{S})\right]
      + n^{-1} E\left[\|\hat{g}\,S-\hat{g}\,\hat{S}\|^2\right] \\
  &=& \underbrace{n^{-1} E\left[\|X-\hat{g}\,S\|^2\right]}_{D_g}
      + \underbrace{n^{-1} E\left[\|\hat{g}\,S-\hat{g}\,\hat{S}\|^2\right]}_{D_s},
\end{eqnarray*}
where the omitted cross term is zero due to the independence of $g$ and $\hat{g}$ from $S$~\cite{HamkinsZ2002}.
The gain distortion, $D_g$, is given by
\[
D_g=\frac{1}{n}\int^{\infty}_0(r-\hat{g}(r))^2f_g(r)\,dr,
\]
where $\hat{g}(\cdot)$ is the quantized gain and $f_g(\cdot)$ is the pdf of $g$.

Conditioned on the gain codeword $\hat{g}_j$ chosen, the shape $S$ is distributed uniformly on the unit sphere in $\R^n$, which has
surface area $S_n=2\pi^{n/2}/\Gamma(n/2)$.
Thus, as shown in~\cite{HamkinsZ2002}, for asymptotically high shape rate $R_s$, the conditional distortion $E[\|S-\hat{S}\|^2 \mid \hat{g_j}]$ is equal to the distortion of the lattice quantizer with codebook $(\Lambda,d_{\Lambda}^j)$ for a uniform source in $\mathbb{R}^{n-1}$. Thus,
\begin{equation}
E\left[\|S-\hat{S}\|^2 \mid \hat{g_j}\right]=(n-1)G(\Lambda)V_j(\Lambda)^{\Frac{2}{(n-1)}},
\label{eq:conditionalDistortion}
\end{equation}
where $V_j(\Lambda)$ is the volume of a Voronoi region of the $(n-1)$-dimensional lattice $(\Lambda,d^j_{\Lambda})$. Therefore, for a given gain codebook $\{\hat{g}_1,\hat{g}_2,\ldots,\hat{g}_J\}$, the shape distortion $D_s$ can be approximated by
\begin{eqnarray*}
 D_s &=& \frac{1}{n}E\left[\|\hat{g}\,S-\hat{g}\,\hat{S}\|^2\right]
      =  \frac{1}{n}\sum_{j=1}^J p_j\,\hat{g}_j^2E\left[\|S-\hat{S}\|^2 \mid \hat{g}=\hat{g}_j\right] \\
    &\approxlabel{a}& \frac{1}{n}\sum_{j=1}^J p_j\,\hat{g}_j^2(n-1)G(\Lambda)V_j(\Lambda)^{\Frac{2}{(n-1)}} \\
    &\approxlabel{b}& \frac{1}{n} \sum_{j=1}^J p_j\,\hat{g}_j^2(n-1)G(\Lambda)\left(\Frac{S_n}{M_j}\right)^{\Frac{2}{(n-1)}} \\
        &=& \frac{n-1}{n}G(\Lambda)S_n^{\Frac{2}{(n-1)}}\sum_{j=1}^J p_j\,\hat{g}_j^2M_j^{-\Frac{2}{(n-1)}}
        =C \cdot \sum_{j=1}^J p_j\,\hat{g}_j^2M_j^{\tfrac{-2}{n-1}},
\end{eqnarray*}
where $p_j$ is the probability of $\hat{g}_j$ being chosen;
(a) follows from (\ref{eq:conditionalDistortion});
(b) follows from the high-rate assumption and neglecting the overlapping regions,
with $M_j$ representing the number of codewords in the shape subcodebook associated with $\hat{g}_j$; and
\begin{equation}
C\defeq \frac{n-1}{n}G(\Lambda){\left(\Frac{2\pi^{n/2}}{\Gamma(n/2)}\right)}^{\Frac{2}{(n-1)}}.\label{eq:constant}
\end{equation}

\subsection{Rate Allocations}
\label{sec:wrapped-rate-alloc}
The optimal rate allocation for high-resolution approximation to WSC given below will be used as the rate allocation across subcodebooks in our CPCs.
\subsubsection{Variable-Rate Coding}
Before stating the theorem, we need the following lemma.
\begin{lem}\label{thm:limit}
If there exist constants $C_s$ and $C_g$ such that
\[
  \lim_{R_s\rightarrow\infty}D_s\cdot 2^{2(\Frac{n}{(n-1)})R_s}=C_s
  \qquad
  \mbox{and}
  \qquad
  \lim_{R_g\rightarrow\infty}D_g\cdot 2^{2nR_g}=C_g\label{eq:lem-limit2},
\]
then the minimum of $D=D_s+D_g$ subject to the constraint $R=R_s+R_g$ satisfies
$$
\lim_{R\rightarrow\infty}D2^{2R}=\frac{n}{(n-1)^{1-\Frac{1}{n}}}\cdot C_g^{\Frac{1}{n}}C_s^{1-\Frac{1}{n}}
$$
and is achieved by $R_s=R_s^{*}$ and $R_g=R_g^{*}$, where
\begin{eqnarray}
     R_s^{*} & = & \left(\frac{n-1}{n}\right)\left[R+\frac{1}{2n}\log\left(\frac{C_s}{C_g}\cdot\frac{1}{n-1}\right)\right], \label{eq:shapeRate} \\
     R_g^{*} & = & \left(\frac{1}{n}\right)\left[R-\frac{n-1}{2n}\log\left(\frac{C_s}{C_g} \cdot \frac{1}{n-1}\right)\right]. \label{eq:gainRate}
\end{eqnarray}
\end{lem}
\begin{IEEEproof}
See~\cite[Thm.~1]{HamkinsZ2002}.
\end{IEEEproof}

\begin{thm}\label{thm:varRate}
    Let ${X}\in\mathbb{R}^n$ be an i.i.d.\ $\mathcal{N}(0,\sigma^2)$ vector, and let $\Lambda$ be a lattice in $\mathbb{R}^{n-1}$ with normalized second moment $G(\Lambda)$. Suppose ${X}$ is quantized by an $n$-dimensional shape--gain VQ at rate $R=R_g+R_s$ with gain-dependent shape codebook constructed from $\Lambda$ with different minimum distances. Also, assume that a variable-rate coding follows the quantization. Then, the asymptotic decay of the minimum mean-squared error $D$ is given by
    \begin{equation}
        \lim_{R\rightarrow\infty}D2^{2R}=\frac{n}{(n-1)^{1-\Frac{1}{n}}}\cdot C_g^{\Frac{1}{n}}C_s^{1-\Frac{1}{n}}
    \end{equation}
    and is achieved by $R_s=R_s^{*}$ and $R_g=R_g^{*}$, where $R_s^{*}$ and $R_g=R_g^{*}$ are given in (\ref{eq:shapeRate}) and (\ref{eq:gainRate}),
    \[
     C_s = \frac{n-1}{n}G(\Lambda){\left(2\pi^{n/2}/\Gamma(n/2)\right)}^{\Frac{2}{(n-1)}} \cdot 2\sigma^2 e^{\psi(n/2)},\quad
     C_g = \sigma^2 \cdot \frac{3^{n/2}\Gamma^3(\frac{n+2}{6})}{8n\Gamma(n/2)},
    \]
     and $\psi(\cdot)$ is the digamma function.
\end{thm}

\begin{IEEEproof}
  We first minimize $D_s$ for a given gain codebook
  $\{\hat{g}_j\}_{j=1}^J$.  From (\ref{eq:constant}),
  ignoring the constant $C$, we must perform the minimization
    \begin{equation}
        \min_{M_1,\ldots,M_J}\ \ts\sum_{j=1}^J p_j \, \hat{g}_j^2 \, M_j^{\Frac{2}{(1-n)}}\quad
        \mbox{subject to} \:         \ts\sum_{j=1}^J p_j\log M_j=nR_s.\label{eq:constraint2}
    \end{equation}
  Using a Lagrange multiplier to get an unconstrained problem,
  we obtain the objective function
  $$
        f=\ts\sum_{j=1}^J p_j \, \hat{g}_j^2 \, M_j^{\Frac{2}{(1-n)}}-\lambda\ts\sum_{j=1}^J p_j\log M_j.
  $$
  Neglecting the integer constraint, we can take the partial derivatives
  $$
        \frac{\partial f}{\partial M_j}=\frac{2}{1-n}p_j\,\hat{g}_j^2M_j^{\Frac{(n+1)}{(1-n)}}-\lambda p_jM_j^{-1},\qquad 1\leq j\leq J.
  $$
  Setting $\frac{\partial f}{\partial M_j}=0,\;1\leq j\leq J$, yields
  \begin{equation}
        M_j=\left[\Frac{\lambda (1-n)}{(2\hat{g}_j^2)}\right]^{\Frac{(1-n)}{2}}\label{eq:Mj}.
  \end{equation}
    Substituting into the constraint (\ref{eq:constraint2}), we get
    \[
        \ts\sum_{j=1}^J p_j\log \left[\Frac{\lambda (1-n)}{(2{g}_j^2)}\right]^{\Frac{(1-n)}{2}}=nR_s.
    \]
    Thus,
    \[
        \left[\Frac{\lambda(1-n)}{2}\right]^{\Frac{(1-n)}{2}} = 2^{nR_s-(n-1)\sum_{k=1}^J p_k\log \hat{g}_k} = 2^{nR_s-(n-1)E[\log\hat{g}]}.
    \]
    Therefore, it follows from (\ref{eq:Mj}) that the optimal size for the $j$th shape subcodebook for a given gain codebook is
    \begin{equation}
        M_j=\hat{g}_j^{n-1}\cdot 2^{nR_s^*-(n-1)E[\log\hat{g}]},\ \  1\leq j\leq J.\label{eq:sizeAllocate1}
    \end{equation}
    The resulting shape distortion is
    \[
        D_s \approx C\cdot \sum_{j=1}^J p_j\,\hat{g}_j^2\left(\hat{g}_j^{n-1}2^{nR_s^*-(n-1)E[\log\hat{g}]}\right)^{\Frac{2}{(1-n)}}
        = C\cdot 2^{2E[\log\hat{g}]}\cdot 2^{-2(\Frac{n}{(n-1)})R_s^*}\,,
    \]
    where $C$ is the same constant as specified in (\ref{eq:constant}). Hence,
    \begin{equation}
        \lim_{R\rightarrow\infty}D_s\cdot 2^{2(\Frac{n}{(n-1)})R_s^*} = C\cdot \lim_{R_g^*\rightarrow\infty}2^{2E[\log\hat{g}]}
        \stackrel{(a)}{=}C\cdot 2^{2E[\log{g}]} \stackrel{(b)}{=}C\cdot 2\sigma^2 e^{\psi(n/2)} =C_s, \label{eq:limit1}
    \end{equation}
    where (a) follows from the high-rate assumption; and (b) follows from computing the expectation $E[\log{g}]$. On the other hand, it is shown in~\cite[Thm.~1]{HamkinsZ2002} that
    \begin{equation}
        \lim_{R\rightarrow\infty}D_g\cdot 2^{2n(R-R_s^*)}=\lim_{R\rightarrow\infty}D_g\cdot 2^{2nR_g^*}=C_g\cdot\label{eq:limit2}
    \end{equation}
    The limits (\ref{eq:limit1}) and (\ref{eq:limit2}) now allow us to apply Lemma~\ref{thm:limit} to obtain the desired result.
\end{IEEEproof}

\begin{figure}
 \begin{center}
  \includegraphics[width=\widthA]{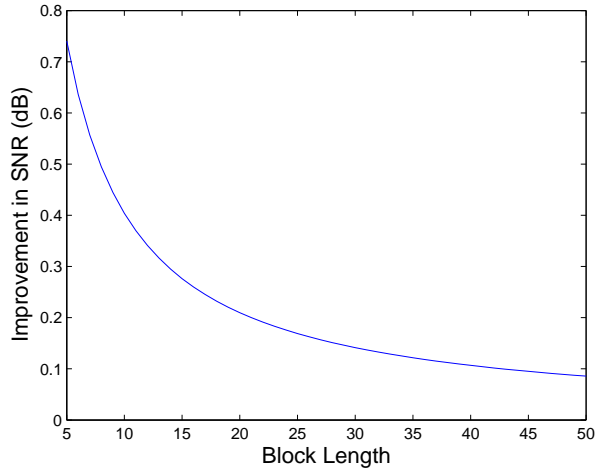}
  \caption{\label{fig:improvedSNR} Improvement in signal-to-quantization noise ratio of WSC with gain-dependent shape quantizer specified in (\ref{eq:improvement}), as compared to the asymptotic rate--distortion performance given in~\cite[Thm.~1]{HamkinsZ2002}}
 \end{center}
\end{figure}

Through this theorem we can verify the rate--distortion improvement as compared to independent shape--gain encoding by comparing $C_g$ and $C_s$ in the distortion formula to the analogous quantities in \cite[Thm.~1]{HamkinsZ2002}. $C_g$ remains the same whereas $C_s$, which plays a more significant role in the distortion formula, is scaled by a factor of $2e^{\psi(n/2)}/n <1$. In particular, the improvement in signal-to-quantization noise ratio achieved by the WSC with gain-dependent shape codebook is given by
\begin{equation}
    \DeltaSNR \,(\mbox{in dB}) =-10(1-\Frac{1}{n})\log_{10}(\Frac{2e^{\psi(n/2)}}{n}). \label{eq:improvement}
\end{equation}
From the theory of the gamma function~\cite[Eq.~29]{Jensen1916}, we know that, for $s\in\mathbb{C}$,
$$\lim_{|s|\rightarrow \infty}[\psi(s)-\ln(s)]=0.$$
It follows that $[\psi(n/2)-\ln(n/2)]\rightarrow 0$, and thus $\DeltaSNR(n)\rightarrow 0$, as $n\rightarrow \infty$; this is not surprising because of the ``sphere hardening'' effect. This improvement is plotted in Figure~\ref{fig:improvedSNR} as a function of block length $n$ in the range between 5 and 50.

\subsubsection{Fixed-Rate Coding}
A similar optimal rate allocation is possible for fixed-rate coding.

\begin{thm}\label{thm:fixedRate}
    Let ${X}\in\mathbb{R}^n$ be an i.i.d.\ $\mathcal{N}(0,\sigma^2)$ vector, and let $\Lambda$ be a lattice in $\mathbb{R}^{n-1}$ with normalized second moment $G(\Lambda)$. Suppose ${X}$ is quantized by an $n$-dimensional shape--gain VQ at rate $R$ with gain-dependent shape codebook constructed from $\Lambda$ with different minimum distances. Also, assume that $J$ gain codewords are used and that a fixed-rate coding follows the quantization. Then, the optimal number of codewords in each subcodebook is
    \begin{equation}
        M_j = 2^{nR}\,\cdot\frac{\left(p_j\hat{g}^2_j\right)^{\Frac{(n-1)}{(n+1)}}}{\sum_{k=1}^J\left(p_k\hat{g}^2_k\right)^{\Frac{(n-1)}{(n+1)}}}, \qquad 1\leq j\leq J, \label{eq:sizeAllocate2}
    \end{equation}
    where $\{\hat{g}_1, \hat{g}_2,\ldots, \hat{g}_J\}$ is the optimal gain codebook. The resulting asymptotic decay of the shape distortion $D_s$ is given by
    \begin{equation}
        \lim_{R\rightarrow\infty}D_s2^{2(\Frac{n}{(n-1)})R}=C \cdot\left[\sum_{j=1}^J (p_j \hat{g}_j^2)^{\tfrac{n-1}{n+1}}\right]^{\tfrac{n+1}{n-1}},
    \end{equation}
    where $p_j$ is probability of $\hat{g}_j$ being chosen and $C$ is the same constant as given in (\ref{eq:constant}).
\end{thm}

\begin{IEEEproof}
  For a given gain codebook $\{\hat{g}_j\}_{j=1}^J$, the optimal subcodebook
  sizes are given by the optimization
  \begin{equation}
    \min_{M_1,\ldots,M_J}\   \ts\sum_{j=1}^J p_j\,\hat{g}_j^2\,M_j^{\Frac{2}{(1-n)}} \quad
    \mbox{subject to} \:    \ts\sum_{j=1}^J M_j=2^{nR}.\label{eq:constraint3}
  \end{equation}
  Similarly to the variable-rate case, we can use a Lagrange multiplier to obtain an unconstrained optimization with the objective function
  $$
    h=\ts\sum_{j=1}^{J}p_j\hat{g}_j^2 M_j^{\Frac{2}{(1-n)}}-\lambda\ts\sum_{j=1}^{J}M_j.
  $$
  Again, assuming high rate, we can ignore the integer constraints on $M_j$ to take partial derivatives. Setting them equal to zero, one can obtain
  \begin{equation}
    M_j=\left[\Frac{\lambda (1-n)}{(2p_j\hat{g}_j^2)}\right]^{\Frac{(1-n)}{(n+1)}}.\label{eq:Mj2}
  \end{equation}
  Substituting into the constraint (\ref{eq:constraint3}) yields
  \[
    \ts\sum_{j=1}^{J} \left[\Frac{\lambda (1-n)}{(2p_j\hat{g}_j^2)}\right]^{\Frac{(1-n)}{(n+1)}}=2^{nR}.
  \]
  Hence,
  \begin{equation}
    \lambda^{\Frac{(n-1)}{(n+1)}} = 2^{-nR}\, \sum_{k=1}^J \left(\frac{1-n}{2p_k\hat{g}_k^2}\right)^{\Frac{(1-n)}{(n+1)}}. \label{eq:lambda}
  \end{equation}
  Combining (\ref{eq:lambda}) and (\ref{eq:Mj2}) give us
  \[
    M_j = \lambda^{\Frac{(1-n)}{(n+1)}}\,\left(\frac{1-n}{2p_j\hat{g}_j^2}\right)^{\Frac{(1-n)}{(n+1)}}
        = 2^{nR} \, \frac{\left(p_j\hat{g}^2_j\right)^{\Frac{(n-1)}{(n+1)}}}{\sum_{k=1}^J\left(p_k\hat{g}^2_k\right)^{\Frac{(n-1)}{(n+1)}}},\qquad 1\leq j\leq J.
  \]
  With the high-rate assumption, the resulting shape distortion will be
  \begin{eqnarray}
    D_s &=& C \, \sum_{j=1}^J p_j\hat{g}_j M_j^{\Frac{2}{(1-n)}}
         =C \, \sum_{j=1}^Jp_j\hat{g}_j\left[\frac{2^{nR}(p_j\hat{g}_j)^{\Frac{(n-1)}{(n+1)}}}{\sum_{k=1}^J(p_k\hat{g}_k^2)^{\Frac{(n-1)}{(n+1)}}}\right]^{\Frac{2}{(1-n)}}\nonumber\\
        &=& C\cdot 2^{-2(\Frac{n}{(n-1)})R}\left[\sum_{j=1}^J (p_j \hat{g}_j^2)^{\Frac{(n-1)}{(n+1)}}\right]^{\frac{n+1}{n-1}}
  \end{eqnarray}
  where $C=\frac{n-1}{n}G(\Lambda){\left(2\pi^{n/2}/\Gamma(n/2)\right)}^{\Frac{2}{(n-1)}}$, completing the proof.
\end{IEEEproof}

\begin{figure}
\centering
\includegraphics[width=\widthA]{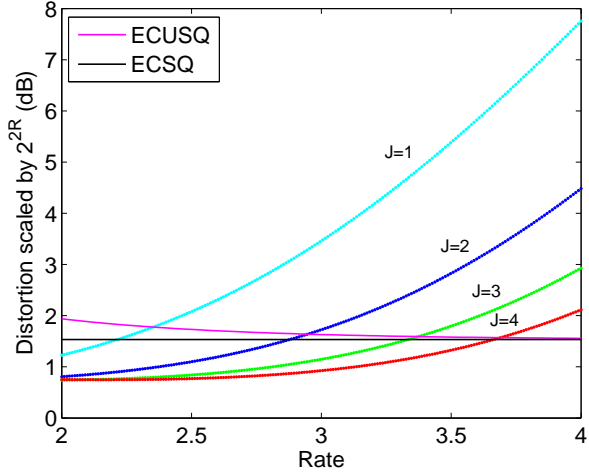}
\caption{\label{fig:fixedRateWSC} High-resolution approximation of the rate--distortion performance of WSC with gain-dependent shape codebooks and fixed-rate coding for an i.i.d.\ $\mathcal{N}(0,1)$ source with block length $n=25$.}
\end{figure}

Figure~\ref{fig:fixedRateWSC} illustrates the resulting performance as a function of the rate
for several values of $J$.  As expected, for a fixed block size $n$, higher rates require higher values of $J$ (more concentric spheres) to attain good performance, and the best performance is improved by increasing the maximum value for $J$.

\subsection{Using WSC Rate Allocation for Permutation Codes}
In this section we use the optimal rate allocations for WSC to guide the
design of
CPCs at a given rate.
The rate allocations are used to set target sizes for each subcodebook.
Then for each subcodebook $\mathcal{C}_j$,
a composition meeting the constraint on $M_j$ is selected
(using heuristics inspired by Conjecture~\ref{conjecture2}).
Algorithm~\ref{alg:lloyd} of Section~\ref{sec:local} is then used for those compositions
to compute the actual rate and distortion.

For the variable-rate case, Theorem~\ref{thm:varRate} provides the key rate allocation step in the design procedure given in Algorithm~\ref{alg:varRate}. Similarly, Theorem~\ref{thm:fixedRate} leads to the design procedure for the fixed-rate case given in Algorithm~\ref{alg:fixedRate}. Each case requires as input not only the rate $R$ but also the number of initial codewords $J$.

\begin{figure}[t]
 \centering
  \includegraphics[width=\widthA]{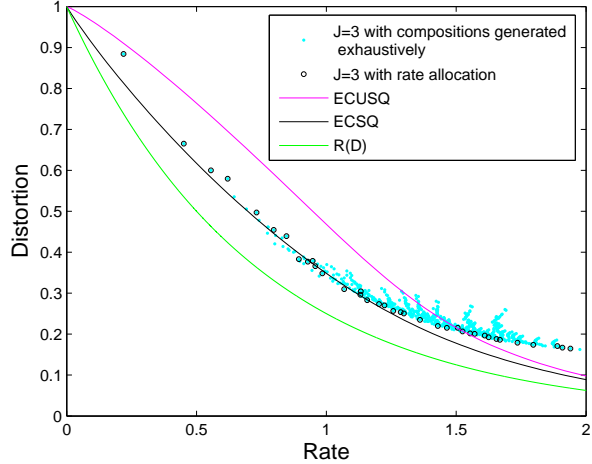}
  \caption{\label{fig:RDapprx}
    Rate--distortion performance of fixed-rate CPCs designed with compositions guided by the WSC rate allocation and Algorithm~\ref{alg:lloyd}, in comparison with codes designed with exhaustive search over a heuristic subset of compositions guided by Conjecture~\ref{conjecture2}. Computation uses i.i.d.\ $\mathcal{N}(0,1)$ source, $n=7$, and $J=3$.}
\end{figure}

\begin{figure}
\centering
\includegraphics[width=\widthA]{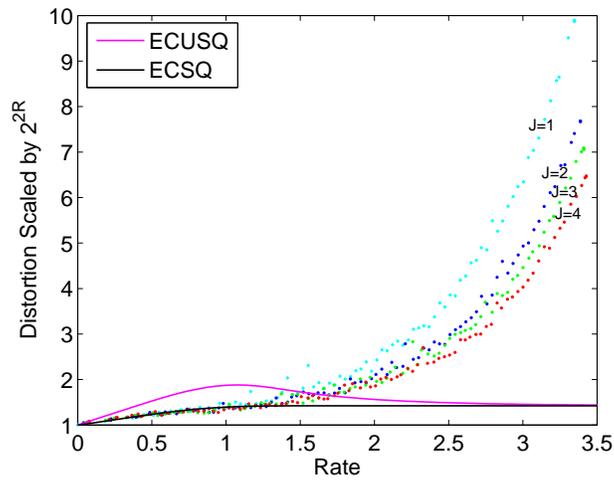}
\caption{\label{fig:fixedRatePC} Rate--distortion performance for variable-rate coding of i.i.d.\
  $\mathcal{N}(0,1)$ source with block length $n=25$.
  CPCs with different compositions are designed using
  rate allocations from Theorem~\ref{thm:fixedRate} and initial codewords locally
  optimized by Algorithm~\ref{alg:lloyd}. The rate allocation computation assumes
  $G(\Lambda_{24})\approx 0.065771$~\cite[p. 61]{ConwayS1998}.}
\end{figure}

Results for the fixed-rate case are plotted in Figure~\ref{fig:RDapprx}.
This demonstrates that using the rate allocation of WSC with gain-dependent shape codebook actually yields good CPCs for most of the rates. Figure~\ref{fig:fixedRatePC} demonstrates the improvement that comes with allowing more initial codewords. The distortion is again computed empirically from Gaussian samples.
It has a qualitative similarity with Figure~\ref{fig:fixedRateWSC}.

\begin{algorithm}
\caption{Design Algorithm for Variable-Rate Case}
\label{alg:varRate}
\begin{tabular}{p{6.5in}}
    \begin{enumerate}
        \item Compute $R^*_s$ and $R^*_g$ from (\ref{eq:shapeRate}) and (\ref{eq:gainRate}), respectively.
        \item For $1\leq j\leq J$, compute $M_j$ from (\ref{eq:sizeAllocate1}).
        \item For $1\leq j\leq J$, search through all possible compositions of $n$ that satisfy Conjecture~\ref{conjecture2}, choosing the one that produces the number of codewords closest to $M_j$.
        \item Run Algorithm~\ref{alg:lloyd} for the $J$ compositions chosen in step 4 to generate the initial codewords and to compute the actual rate and distortion.
    \end{enumerate}\\
\end{tabular}
\end{algorithm}

\begin{algorithm}
\caption{Design Algorithm for Fixed-Rate Case}
\label{alg:fixedRate}
\begin{tabular}{p{6.5in}}
        \begin{enumerate}
        \item Use the scalar Lloyd-Max algorithm to optimize $J$ gain codewords.
        \item For $1\leq j\leq J$, compute $M_j$ from (\ref{eq:sizeAllocate2})
        \item Repeat steps 3 and 4 of Algorithm~\ref{alg:varRate}.
    \end{enumerate}\\
\end{tabular}
\end{algorithm}

\section{Conclusions}
\label{sec:conclusions}
We have studied a generalization of permutation codes in which more than
one initial codeword is allowed.
This improves rate--distortion performance while adding very little to
encoding complexity.
However, the design complexity is increased considerably.
To reduce the design complexity, we explore a method introduced by Lu \emph{et al.} of restricting the
subcodebooks to share a common composition; and we introduce a method of allocating rates across
subcodebooks using high-resolution analysis of wrapped spherical codes.
Simulations suggest that these heuristics are effective,
but obtaining theoretical guarantees remains an open problem.

\section*{Acknowledgment}
The authors thank Danilo Silva for bringing the previous work on composite permutation codes to their attention.

\appendices
\section{Proof of Proposition~\ref{thm:reverse}}
\label{app:proof}
Consider a new composition $\{n_1',n_2',\ldots,n_K'\}$ obtained by swapping $n_m$ and $n_{m+1}$, i.e.,
     \[
        n_i'=\left\{ \begin{array}{ll}
        n_i,     & i \neq m \mbox{  or  } m+1; \\
        n_{m+1}, & i=m; \\
        n_m,     & i=m+1.
        \end{array} \right.
    \]
    Let $\{I_i'\}$ denote groups of indices generated by composition $\{n_i'\}$.
    Suppose that $D$ is the optimal distortion associated with $\{n_i\}$,
    \[
        D= n^{-1} E\left[\min_{1\leq j \leq J}\ts\sum_{i=1}^K \sum_{\ell \in I_i}\left(\eta_\ell-\mu^j_i \right)^2\right],
    \]
    where $\{\mu^j_i\}$ is the optimum of the minimization of the right side over $\Omega_m$.
    Consider a suboptimal distortion $D'$ associated with $\{n_i'\}$,
    \[
        D'= n^{-1} E\left[\min_{1\leq j \leq J}\ts\sum_{i=1}^K \sum_{\ell \in I'_i}\left(\eta_\ell-\tilde{\mu}^j_i \right)^2\right],
    \]
    where $\{\tilde{\mu}_i^j\}$ is constructed from $\{\mu^j_i\}$ as follows: for each $j$,
    \begin{equation}
        \tilde{\mu}^j_i=\left\{ \begin{array}{ll}
        \mu^j_i, & i \neq m \mbox{  or  } m+1; \\
        \frac{2n_m\mu^j_m+(n_{m+1}-n_m)\mu^j_{m+1}}{n_m+n_{m+1}}, & i=m; \\
        \frac{(n_{m}-n_{m+1})\mu^j_m+ 2n_{m+1}\mu^j_{m+1}}{n_m+n_{m+1}}, & i=m+1.
        \end{array} \right. \label{eq:construction}
    \end{equation}

    Note that, for the above construction, we have
    $\tilde{\mu}^j_m-\tilde{\mu}^j_{m+1}=\mu^j_m-\mu^j_{m+1}$, for all $j \in \{1,2,\ldots ,J\}$.
    Therefore $\{\tilde{\mu}^j_i\}$ also satisfies $\Omega_m$, and so forms a valid codebook corresponding to composition $\{n'_i\}$. Thus, it will be sufficient if we can show $D>D'$. On the other hand, it is easy to verify that, for all $j$,
    \[
        n_{m+1}(\tilde{\mu}^j_m)^2+n_{m}(\tilde{\mu}^j_{m+1})^2 = n_{m}(\mu^j_m)^2 + n_{m+1}(\mu^j_{m+1})^2 .
    \]
    Hence,
    \begin{equation}
        \sum_{i=1}^K n_i'(\tilde{\mu}^j_i)^2= \sum_{i=1}^K n_i(\mu^j_i)^2 \;, \;\;\; \mbox{for all $j$}. \label{eq:sum}
    \end{equation}

    Now consider the difference between $D$ and $D'$:
    \begin{eqnarray*}
    \Delta &=& n(D-D')
        = E\left[\min_{j}\sum_{i=1}^K \sum_{\ell \in I_i}\left(\eta_\ell-\mu^j_i \right)^2\!\!-\min_{j}\sum_{i=1}^K \sum_{\ell \in I_i'}\left(\eta_\ell-\tilde{\mu}^j_i \right)^2\right] \\
        &\geqlabel{a}& E\left[\min_{j}\left\{ \sum_{i=1}^K\left(n_i(\mu^j_i)^2-2\mu^j_i\sum_{\ell\in I_i}\eta_\ell\right)-\sum_{i=1}^K\left(n_i'(\tilde{\mu}^j_i)^2-2\tilde{\mu}^j_i\sum_{\ell\in I_i'}\eta_\ell\right)\right\}\right] \\
        &\eqlabel{b}& 2E\left[\min_{j}\left\{\tilde{\mu}^j_m\sum_{\ell=L+1}^{L+r}\eta_\ell+\tilde{\mu}^j_{m+1}\sum_{\ell=L+r+1}^{L+r+q}\eta_\ell -\mu^j_m\sum_{\ell=L+1}^{L+q}\eta_\ell-\mu^j_{m+1}\sum_{\ell=L+q+1}^{L+q+r}\eta_\ell\right\}\right],
    \end{eqnarray*}
    where (a) uses the fact that $\min{f}-\min{g} \geq \min\{f-g\}$, for arbitrary functions $f,g$;
    and (b) follows from (\ref{eq:sum}) in which $q=n_m$, $ r=n_{m+1}$, and $L=n_1+n_2+\cdots + n_{m-1}$. Now using the formulae of $\tilde{\mu}^j_{m}$ and $\tilde{\mu}^j_{m+1}$ in (\ref{eq:construction}), we obtain
    \begin{eqnarray*}
    \Delta& \geq& 2E\left[ \min_{j} \left\{\frac{(q-r)(\mu^j_{m}-\mu^j_{m+1})}{q+r}\sum_{\ell=L+1}^{L+r}\eta_\ell\right.\right.
	  -\frac{2r(\mu^j_{m}-\mu^j_{m+1})}{q+r}\sum_{\ell=L+r+1}^{L+q}\eta_\ell\\
        & & \hspace{0.4in} \left.\left.+\frac{(q-r)(\mu^j_{m}-\mu^j_{m+1})}{q+r}\sum_{\ell=L+q+1}^{L+q+r}\eta_\ell\right\}\right]\\
        &=&  \frac{2r(q-r)}{q+r}E\left[\min_{j}\left\{(\mu^j_m-\mu^j_{m+1})\zeta\right\}\right] \\
        &\eqlabel{a}&  \frac{2r(q-r)}{q+r}\left[\overline{\zeta}_{+}\cdot\min_{j}\left\{\mu^j_m-\mu^j_{m+1}\right\}-\overline{\zeta}_{-}\cdot\max_{j}\left\{\mu^j_m-\mu^j_{m+1}\right\}\right]\\
        &\geqlabel{b}&  0,
    \end{eqnarray*}
    where $\zeta$ is the random variable specified in (\ref{eq:rv}); (a) follows from the total expectation theorem; and
    (b) follows from constraint $\Omega_m$ and that $q>r$. The nonnegativity of $\Delta$ has proved the proposition.

\section{The Number and Density of Distinct Rates}
\label{app:rate-points}
In this appendix, we discuss the distinct rate points
at which fixed-rate ordinary PCs, CPCs with common compositionss,
and CPCs with possibly-different compositionss may operate.  For brevity,
we restrict attention to Variant~I codes.

The number of codewords (and therefore the rate) for an ordinary
PC is determined by the multinomial coefficient
(\ref{eq:multinomial}).
The multinomial coefficient is invariant to the order of the $n_i$s,
and so we are interested in the number of unordered compositionss (or \emph{integer partitions}) of
$n$, $P(n)$.  Hardy and Ramanujan~\cite{HardyR1918} gave the asymptotic formula:
$
P(n) \sim \tfrac{e^{\pi\sqrt{2n/3}}}{4n\sqrt 3}
$.
One might think that the number of possible distinct rate points is $P(n)$,
but different sets of $\{n_i\}$ can yield the same multinomial coefficient.
For example, at $n = 7$ both $(3,2,2)$ and $(4,1,1,1)$ yield $M = 210$.
Thus we are instead interested in the number of distinct multinomial coefficients, $N_{\rm{mult}}(n)$
\cite{AndrewsKZ2006}, \cite[A070289]{Sloane}.  Clearly $P(n) \ge N_{\rm{mult}}(n)$.
A lower bound to $N_{\rm{mult}}(n)$ is the number of unordered partitions of $n$ into
parts that are prime, $P_{\mathbb{P}}(n)$, with asymptotic formula:
$
P_{\mathbb{P}}(n) \sim \exp\left\{\frac{2\pi\sqrt n}{\sqrt{3\log n}}\right\}
$.
Thus the number of distinct rate points for ordinary PCs
grows exponentially with block length.

It follows easily that the average density of distinct rate points on the
interval of possible rates grows without bound.  Denote this average density
by $\delta(n)$.  The interval of possible rates is $[0, \log n!/n]$, so
applying the upper and lower bounds gives the asymptotic expression
\[
\frac{n\exp\left\{\frac{2\pi}{\sqrt 3}\sqrt{\frac{n}{\log n}}\right\}} {\log n!}
  \lesssim \delta(n)
  \lesssim \frac{e^{\pi\sqrt{2n/3}}}{4\sqrt 3 \log n!} \mbox{.}
\]
Taking the limits of the bounds then yields
$
\lim_{n\to\infty} \delta(n) =  +\infty$.

The following proposition addresses the maximum gap between rate points,
giving a result stronger than the statement on average density.
\begin{prop}
The maximum spacing between any pair of rate points goes to $0$ as $n \to \infty$.
\end{prop}
\begin{IEEEproof}
First note that there are rate points at
$
0,\, \tfrac{\log[n]}{n},\, \tfrac{\log[(n)(n-1)]}{n},\, \tfrac{\log[(n)(n-1)(n-2)]}{n},\, \cdots
$
induced by integer partitions
$
(n),\, (n-1,1),\, (n-2,1,1),\, (n-3,1,1,1),\, \ldots
$
The lengths of the intervals between these rate points is
$
\tfrac{\log(n)}{n},\, \tfrac{\log(n-1)}{n},\, \tfrac{\log(n-2)}{n},\, \ldots
$
which decreases as one moves to larger rate points, so the first one is the largest.
If there are other achievable rates between $\frac{\log[n]}{n}$ and $\frac{\log[(n)(n-1)]}{n}$ and so on, they
only act to decrease the size of the intervals between successive rate points.
So the interval between $0$ and $\frac{\log[n]}{n}$ is the largest.

Taking $n \to\infty$ for the largest interval gives
$\lim_{n \to\infty} \tfrac{\log n}{n} = 0$,
so the maximum distance between any rate points goes to zero.
\end{IEEEproof}

$N_{\rm{mult}}(n)$ is the number of distinct rate points
for ordinary PCs.  If fixed-rate CPCs
are restricted to have common compositions, then they too have
the same number of distinct rate points.  If different compositions
are allowed, the number of distinct rate points may increase dramatically.

Recall the rate expression (\ref{eq:fixed-rateRate}), and notice that
distinct values of $\sum M_j$ will yield distinct rate points.  Somewhat similarly to
the distinct subset sum problem \cite[pp.~174--175]{Guy2004}, we want to
see how many distinct sums are obtainable from subsets of size $J$ selected
with replacement from the possible multinomial coefficients of a given block length $n$.
This set is denoted $\mathcal{M}(n)$ and satisfies $|\mathcal{M}(n)| = N_{\rm{mult}}(n)$;
for example, $\mathcal{M}(4) = \{1,4,6,12,24\}$.

For a general set of integers of size $N_{\rm{mult}}(n)$, the number
of distinct subset sums is upper-bounded by $\binom{N_{\rm{mult}}(n) + J - 1}{J}$.
This is achieved, for example, by the set $\{1^1,2^2,\ldots,N_{\rm{mult}}(n)^{N_{\rm{mult}}(n)}\}$.
The number of distinct subset sums, however, can be much smaller.
For example, for the set $\{1,2,\ldots,N_{\rm{mult}}(n)\}$,
this number is $J \, N_{\rm{mult}}(n) - J + 1$.
We have been unable to obtain a general expression for the set $\mathcal{M}(n)$;
this seems to be a difficult number theoretic problem.  It can be noted, however,
that this number may be much larger than $N_{\rm{mult}}(n)$.

Exact computations for the number of distinct rate points at
small values of $n$ and $J$ are provided in Table~\ref{tab:numPoints}.

\begin{table}
\caption{Number of Rate Points}
\label{tab:numPoints}
\centering
\begin{tabular}{rrrrr}
\hline
$n$ & $J=1$& $J=2$ & $J=3$  & $J=4$ \\ \hline 
$2$ & $2$  & $3$   & $4$    & $5$     \\
$3$ & $3$  & $6$   & $10$   & $15$    \\
$4$ & $5$  & $15$  & $33$   & $56$    \\
$5$ & $7$  & $27$  & $68$   & $132$   \\
$6$ & $11$ & $60$  & $207$  & $517$   \\
$7$ & $14$ & $97$  & $415$  & $1202$  \\
$8$ & $20$ & $186$ & $1038$ & $3888$  \\
$9$ & $27$ & $335$ & $2440$ & $11911$ \\ \hline
\end{tabular}
\end{table}

\bibliographystyle{IEEEtran}
\bibliography{abrv,conf_abrv,lrv_lib}

\begin{thebibliography}{10}
\providecommand{\url}[1]{#1}
\csname url@samestyle\endcsname
\providecommand{\newblock}{\relax}
\providecommand{\bibinfo}[2]{#2}
\providecommand{\BIBentrySTDinterwordspacing}{\spaceskip=0pt\relax}
\providecommand{\BIBentryALTinterwordstretchfactor}{4}
\providecommand{\BIBentryALTinterwordspacing}{\spaceskip=\fontdimen2\font plus
\BIBentryALTinterwordstretchfactor\fontdimen3\font minus
  \fontdimen4\font\relax}
\providecommand{\BIBforeignlanguage}[2]{{%
\expandafter\ifx\csname l@#1\endcsname\relax
\typeout{** WARNING: IEEEtran.bst: No hyphenation pattern has been}%
\typeout{** loaded for the language `#1'. Using the pattern for}%
\typeout{** the default language instead.}%
\else
\language=\csname l@#1\endcsname
\fi
#2}}
\providecommand{\BIBdecl}{\relax}
\BIBdecl

\bibitem{Dunn1965}
J.~G. Dunn, ``Coding for continuous sources and channels,'' Ph.D. dissertation,
  Columbia Univ., New York, May 1965.

\bibitem{BergerJW1972}
T.~Berger, F.~Jelinek, and J.~K. Wolf, ``Permutation codes for sources,''
  \emph{{IEEE} Trans. Inform. Theory}, vol. IT-18, no.~1, pp. 160--169, Jan.
  1972.

\bibitem{Berger1972}
T.~Berger, ``Optimum quantizers and permutation codes,'' \emph{{IEEE} Trans.
  Inform. Theory}, vol. IT-18, no.~6, pp. 759--765, Nov. 1972.

\bibitem{Sakrison1968}
D.~J. Sakrison, ``A geometric treatment of the source encoding of a {G}aussian
  random variable,'' \emph{{IEEE} Trans. Inform. Theory}, vol. IT-14, no.~3,
  pp. 481--486, May 1968.

\bibitem{Jelinek1968}
F.~Jelinek, ``Buffer overflow in variable length coding of fixed rate
  sources,'' \emph{{IEEE} Trans. Inform. Theory}, vol. IT-14, no.~3, pp.
  490--501, May 1968.

\bibitem{GoyalSW2001}
V.~K. Goyal, S.~A. Savari, and W.~Wang, ``On optimal permutation codes,''
  \emph{{IEEE} Trans. Inform. Theory}, vol.~47, no.~7, pp. 2961--2971, Nov.
  2001.

\bibitem{HamkinsZ2002}
J.~Hamkins and K.~Zeger, ``{G}aussian source coding with spherical codes,''
  \emph{{IEEE} Trans. Inform. Theory}, vol.~48, no.~11, pp. 2980--2989, Nov.
  2002.

\bibitem{Slepian1968}
D.~Slepian, ``Group codes for the {G}aussian channel,'' \emph{Bell Syst. Tech.
  J.}, vol.~47, pp. 575--602, Apr. 1968.

\bibitem{Bona2006}
M.~{B\'{o}na}, \emph{A Walk Through Combinatorics: An Introduction to
  Enumeration and Graph Theory}, 2nd~ed.\hskip 1em plus 0.5em minus 0.4em\relax
  World Scientific Publishing Company, 2006.

\bibitem{LuCG1986}
L.~Lu, G.~Cohen, and P.~Godlewski, ``Composite permutation coding of speech
  waveforms,'' in \emph{Proc. IEEE Int. Conf. Acoust., Speech, Signal Process.
  (ICASSP 1986)}, Apr. 1986, pp. 2359--2362.

\bibitem{LuCG1994}
------, ``Composite permutation codes for vector quantization,'' in \emph{Proc.
  1994 IEEE Int. Symp. Inf. Theory}, Jun. 1994, p. 237.

\bibitem{AbeKN2007}
S.~Abe, K.~Kikuiri, and N.~Naka, ``Composite permutation coding with simple
  indexing for speech/audio codecs,'' in \emph{Proc. IEEE Int. Conf. Acoust.,
  Speech, Signal Process. (ICASSP 2007)}, vol.~4, Apr. 2007, pp. 1093--1096.

\bibitem{FinamoreBS2004}
W.~A. Finamore, S.~V.~B. Bruno, and D.~Silva, ``Vector permutation encoding for
  the uniform sources,'' in \emph{Proc. Data Compression Conf. (DCC 2004)},
  Mar. 2004, p. 539.

\bibitem{Kuhn1955}
H.~W. Kuhn, ``The {H}ungarian method for the assignment problem,'' \emph{Nav.
  Res. Logist. Q.}, vol.~2, no. 1-2, pp. 83--97, Mar. 1955.

\bibitem{Slepian1965}
D.~Slepian, ``Permutation modulation,'' \emph{Proc. {IEEE}}, vol.~53, no.~3,
  pp. 228--236, Mar. 1965.

\bibitem{Berger1982}
T.~Berger, ``Minimum entropy quantizers and permutation codes,'' \emph{{IEEE}
  Trans. Inform. Theory}, vol. IT-28, no.~2, pp. 149--157, Mar. 1982.

\bibitem{DavidN2003}
H.~A. David and H.~N. Nagaraja, \emph{Order Statistics}, 3rd~ed.\hskip 1em plus
  0.5em minus 0.4em\relax Hoboken, NJ: Wiley-Interscience, 2003.

\bibitem{Lloyd1957}
S.~P. Lloyd, ``Least squares quantization in {PCM},'' Jul. 1957, unpublished
  Bell Laboratories Technical Note.

\bibitem{Max1960}
J.~Max, ``Quantizing for minimum distortion,'' \emph{{IRE} Trans. Inform.
  Theory}, vol. IT-6, no.~1, pp. 7--12, Mar. 1960.

\bibitem{MarcoN2006}
D.~Marco and D.~L. Neuhoff, ``Low-resolution scalar quantization for {G}aussian
  sources and squared error,'' \emph{{IEEE} Trans. Inform. Theory}, vol.~52,
  no.~4, pp. 1689--1697, Apr. 2006.

\bibitem{SwaszekT1982}
P.~F. Swaszek and J.~B. Thomas, ``Optimal circularly symmetric quantizers,''
  \emph{J. Franklin Inst.}, vol. 313, no.~6, pp. 373--384, Jun. 1982.

\bibitem{Hamkins1996}
J.~Hamkins, ``Design and analysis of spherical codes,'' Ph.D. dissertation,
  Univ. Illinois at Urbana-Champaign, Sep. 1996.

\bibitem{Jensen1916}
J.~L.~W.~V. Jensen, ``An elementary exposition of the theory of the gamma
  function,'' \emph{Ann. Math.}, vol.~17, no.~3, pp. 124--166, Mar. 1916.

\bibitem{ConwayS1998}
J.~H. Conway and N.~J.~A. Sloane, \emph{Sphere Packings, Lattices, and
  Groups}.\hskip 1em plus 0.5em minus 0.4em\relax New York: Springer-Verlag,
  1998.

\bibitem{HardyR1918}
G.~H. Hardy and S.~Ramanujan, ``Asymptotic formulae in combinatory analysis,''
  \emph{Proc. Lond. Math. Soc.}, vol.~17, no.~1, pp. 75--115, 1918.

\bibitem{AndrewsKZ2006}
G.~E. Andrews, A.~Knopfmacher, and B.~Zimmermann, ``On the number of distinct
  multinomial coefficients,'' \emph{J. Number Theory}, vol. 118, no.~1, pp.
  15--30, May 2006.

\bibitem{Sloane}
\BIBentryALTinterwordspacing
N.~J.~A. Sloane, ``The on-line encyclopedia of integer sequences.'' [Online].
  Available: \url{{http://www.research.att.com/~njas/sequences}}
\BIBentrySTDinterwordspacing

\bibitem{Guy2004}
R.~K. Guy, \emph{Unsolved Problems in Number Theory}, 3rd~ed.\hskip 1em plus
  0.5em minus 0.4em\relax New York: Springer, 2004.

\end{thebibliography}

\end{document}